\input{aipcheck}
\documentclass[
draft            
  ]
  {aipproc}

\layoutstyle{6x9}

\begin{document}

\title{Searching for GRB remnants in nearby galaxies}

\author{S. G. Bhargavi}{
  address={Indian Institute of Astrophysics, Sarjapur Road, Bangalore 560 034 India}
}

\author{J. Rhoads}{
  address={Space Telescope Science Institute, 3700 San Martin Drive,
     Baltimore, MD 21218, USA
}
}

\author{R. Perna}{
  address={Dept. of Astrophysical Sciences, Princeton University, 
4 Ivy Lane, Princeton, NJ, 08544, USA
}
}

\author{J. Feldmeier}{
  address={Case Western Reserve University, Cleveland, OH 44106-1712, USA
}
}
\author{J. Greiner}{
  address={Max-Planck Institute for Extraterrestrial Physics, Munich, Germany
}
}

\begin{abstract}
Gamma Ray Bursts (GRBs) are expected to leave behind GRB remnants,
similar to how ``standard'' supernovae (SN) leave behind SN remnants.
The identification of
these remnants in our own and in nearby galaxies would allow a much closer
look at GRB birth sites, and possibly lead to the discovery of the 
compact object left behind. It would also provide independent constraints
on GRB rates and energetics. We have initiated an observational program to
search for GRB remnants in nearby galaxies. The identification is based on
specific line ratios, such as OIII/$H_{\beta}$ and HeII/$H_{\beta}$, which are 
expected to be unusually high in case of GRB remnants according to 
the theoretical predictions of Perna et al. (2000). The observing strategies and
preliminary studies from a test run at 2.34 m VBT as well as archival data
from planetary nebulae surveys of spiral galaxies are discussed.
\end{abstract}

\maketitle

\section{ Introduction}

The intense X-ray/UV radiation accompanying GRBs has dramatic effects
on their environment: it can photoionize regions of $\sim 100$ pc size (Perna \& Loeb 1998),
and destroy dust on scales of tens of pc (e.g. Waxman \& Draine 2000). 
Moreover, similarly to what happens for SN remnants, a powerful blast wave
is driven into the medium. While it takes a very short time (compared to the
duration of the most intense X-ray UV radiation from the burst) to alter
the equilibrium state of the medium, it takes a very long time for
the medium to recover its original state, as detailed in the next section.
This means that, while it is highly unlikely to observe a GRB in a nearby
galaxy, there is a significant probability to find a GRB remnant.

Identification of GRB remnants in our own and nearby galaxies would allow
a close study of GRB birth sites, and therefore provide independent diagnostic
of their progenitors. Similar to how pulsars are found in association with SN
remnants, the compact remnant objects left over from the GRB explosions 
can then be found in association with GRB remnants. Moreover, an estimate
of the number of GRB remnants in the local universe would allow independent
constraints on GRB rates and energetics.  

\section{GRB Remnants}
Two phases can be distinguished in the life of a GRB remnant:

\begin{description}
\item[Cooling remnants] (Perna, Raymond \& Loeb 2000): due to the radiation flux of the GRB and its afterglow.

 The X-ray/UV radiation accompanying a GRB heats and ionizes the surrounding medium.
 An emission spectrum is expected to be produced from the cooling ionized gas,
  the cooling time being of the order of $\sim 10^5 (T/10^5{\rm K})/(n/{\rm cm^{-3}})$) yr.

\item[Slowing remnants] (Loeb \& Perna 1998; Efremov et al. 1998): due to the slowing blast-wave.

The relativistically expanding blast wave resulting from a GRB explosion
takes $\sim 10^7$ years to slow down and merge with the ISM.

\end{description}
Combining these time scales
with the present GRB rate \\
of $\sim (10^6-10^7 {f_b}\,{\rm yr})^{-1}$ per galaxy ($f_b$ is the beaming fraction)
(e.g. Wijers {\it et al.} 1998),
 it can be seen that there is a substantial probability of finding GRB remnants
in any galaxy at any given time. 

\subsection{Identifying GRB Remnants}

In this paper, we report on our initial search strategy for cooling GRB remnants. 
Although the duration of the cooling phase is much shorter than the 
lifetime of the blastwave, these cooling remnants are much easier to identify due to their
unique spectral signatures that allow one to
distinguish them from other  sources such as shock heated gas in SN remnants, 
HII regions and planetary nebulae (Perna, Raymond \& Loeb 2000).
\begin{itemize}
\item High value for the line ratio [OIII]$\lambda$5007 /$H_{\beta}$ ($\sim 100$ for most of the
cooling phase).
\item Unusally high value for He II $\lambda$4686 /$H_{\beta}$ ratio (up to $\sim 100$ at the
beginning of the cooling phase).
\item Time-dependent increase in the ratio [OIII]/[OII] indicating 
cooling of the gas.
\item High SII$\lambda$6717/$H_{\beta}$ as compared to HII regions.
\end{itemize}

HII regions are also photo-ionized like GRB remnants and can sometimes 
have high OIII$\lambda$5007/$H_{\beta}$
ratios ($\approx$ 3), but are characterized by lower temperatures in comparion
to GRB remnants. Therefore it is useful to measure 
the [OIII]$\lambda$ 5007/[OIII]$\lambda$ 4363
line ratio which is temperature sensitive and increases with time 
in a cooling gas.
The oxygen-rich SNR might show high OIII/$H_{\beta}$ occationally
but only during a brief period of incomplete cooling.
The He II $\lambda$4686 /$H_{\beta}$ ratio is weak in 
both HII regions as well as SNRs.
Further, GRB remnants have physical sizes of $\sim$ 100 pc, 
and can be distinguished
from planetary nebulae (PNe) which look like point sources in external galaxies.


\section{Observations}

\subsection{VBO data}

We have initiated our observational search for cooling GRB remnants 
using the 2.34 m Vainu Bappu Telescope (VBT)
at Kavalur, India,  and plan to use the
new 2 meter Himalayan Chandra Telescope of the Indian Astronomical
Observatory (IOA) at Hanle, featured by cloudless skies and low 
atmospheric water vapour.
Nearby galaxies will be observed in narrow-band filters [OIII]$\lambda$ 5007,
[OIII]$\lambda$ 4363, He II $\lambda$4686 and $H_{\beta}$ to measure the various
line ratios and to identify the candidate GRB remnants for further investigations.

In a test run of observations this summer, NGC 3627 and NGC 3351
were imaged at the 2.34 m VBT using the narrow-band filters
[OIII]$\lambda$ 5007, [OIII]$\lambda$ 4363 and $H_{\beta}$.

\subsection{KPNO archival data}

In addition to the narrow-band data taken with the VBT, we are
searching for GRB remnants using archival data originally taken
to search for planetary nebulae in spiral galaxies (Feldmeier et al.
1997; Ciardullo et al. 2002).  The data consists of narrow band
[O~III] $\lambda$5007 and H$\alpha$ + [NII] exposures, along
with a $\lambda$5300 continuum image.  In some cases, there is
additional $R$ data as well.  The seven galaxies (NGC 891,
2403, 3627, 3351, 3368, 5194/5, 5457) are all luminous spiral
galaxies, and should provide reasonable targets for search.

While the [OIII]$\lambda$5007/$H_{\alpha}$ ratios are typically 5 in PNe, we expect it to be
$\sim$ 30 for a candidate GRB remnant (using $H_{\alpha}$/$H_{\beta}$ of 2.8
for emission nebulae; Osterbrock 1989).

Table~1 shows the telescope parameters.

\begin{table}
\begin{tabular}{lrrr}
\hline
  & \tablehead{1}{r}{b}{VBO}
  & \tablehead{1}{r}{b}{IOA}
  & \tablehead{1}{r}{b}{KPNO} \\ 
\hline
Size & 2.34 m & 2.01 m & 4 m\\
Longitude: & $78^{\circ}49^{\prime} 36^{\prime\prime}$E & 
$78^{\circ}57^{\prime}51^{\prime\prime}$E &
$111^{\circ}36^{\prime}59^{\prime\prime}$W\\
latitude:  & $12^{\circ}34^{\prime} 36^{\prime\prime}$N & 
$32^{\circ}46^{\prime}46^{\prime\prime}$N &
$31^{\circ}57^{\prime}12^{\prime\prime}$N\\
Altitude:  & 725 m & 4500 m & 2100 m\\
Seeing (typ.): & $2^{\prime\prime}.5$ & $< 1^{\prime\prime}$ & $ 1^{\prime\prime}$\\
F-ratio:   & f/3.25 prime  & f/9 cassegrain & f/3.1 prime \\
Image scale: & $0^{\prime\prime}.6$/pix & $0^{\prime\prime}.17$/pix  & 
$0^{\prime\prime}.42$/pix \\
FOV:       & $10^{\prime}\times 10^{\prime}$ & $7^{\prime}\times 7^{\prime}$ &
$14^{\prime}\times 14^{\prime}$\\
\hline
\end{tabular}
\caption{Telescope Parameters}
\label{tab:a}
\end{table}

\section{Search strategies }

The images are reduced in the standard manner using the IRAF software.
All images are aligned and positionally registered before continuum subtraction.
Our goal is to identify candidate GRB remnants, and
follow them up with additional imaging and spectroscopic observations.
Since we have just begun our search, our results are still
preliminary.  In order to separate the potential
GRB remnant candidates from other sources (HII regions, SN remnants,
and PNe), we are using the following selection criteria:

\noindent
1. Objects must have a signal-to-noise greater than 9.\\
2. Objects must appear non-stellar, and have a SExtractor star/galaxy
classifier value less than 0.95. \\
3. Objects must have an [OIII]$\lambda$5007/ H$\alpha$ ratio that is 2$\sigma$
larger than the mean of the distribution.  This removes almost all
H~II regions from the sample, as they have low [O~III]$\lambda$5007 / H$\alpha$
ratios.

Each candidate is then visually inspected to confirm their candidature
against artifacts.  Currently, we are finding tens of
candidates in each galaxy, though there is signifigant scatter
from galaxy to galaxy (NGC 891 having none, most likely due to its edge-on
orientation).  We require further observations to confirm
the list of candidates. 

\section{Conclusions}

In this paper we report the preliminary studies we carried out to
search for the GRB remnants in nearby galaxies
from a test run at VBT as well as archival data from KPNO.
In the present investigations we find about 20-30 candidates in each galaxy.
Additional observations are required to measure other line ratios
and to check the validity of candidates.
We also plan to use data from other existing surveys to find preliminary 
candidates in nearby galaxies.
A photoionized remnant of radius $\sim$ 100 pc would subtend an angle
of $2^{\prime\prime}$ on the sky at the distance of Virgo cluster ($\sim$ 20 Mpc),
where a typical galaxy would subtend 2 arcmin. We require a 
multiple-fibre spectrograph 
 which can take simultaneous spectra across the
entire image of a nearby galaxy. 

\vspace{-0.1in}
\begin{theacknowledgments}
We thank Prof. A. Saha (KPNO) for lending the narrow-band
filters for the observations at VBT.
SGB acknowledges the science visit hosted by STScI and
also Prof. D Lamb, University of Chicago for guest user facility
to carry out a part of the work.

\end{theacknowledgments}



\vspace{-0.2in}
\begin{thebibliography}{}
\bibitem{} Bhargavi, S. G., Cowsik, R. \& Perna, R. 2002, In `GRBs in the afterglow
era',
To appear in Proceedings of 3${\rm rd}$ Rome workshop, Sept 2002  
\bibitem{} Ciardullo, R., Feldmeier, J. J., Jacoby, G. H., Kuzio de Naray, R., Laych
ak, M. B. \& Durrell, P. R. 2002, ApJ, 577, 31
\bibitem{} Efremov, Y. N., Elmegreen, B. G., hodge, P. W. 1998, ApJ, 501L, 163
\bibitem{} Feldmeier, J. J., Ciardullo, R. \& Jacoby, G. H. 1997, ApJ, 479, 231 
\bibitem{} Loeb, A. \& Perna, R. 1998, ApJ, 503L, 35
\bibitem{} Perna, \& Loeb, A. 1998, ApJ, 501, 467
\bibitem{} Perna, R., Raymond, J. \& Loeb, A.  2000 ApJ, 533, 658 
\bibitem{} Waxman, E. \& Draine, B. 2000, ApJ, 537, 796 
\bibitem{} Wijers, R. A. M. J., Bloom, J. S., Bagla, J. S. \& Natarajan. P. 1998, MN
RAS, 294, l13
\end{thebibliography}


\end{document}